# Nodal Knotted Finite-Energy Spatiotemporally Localized Scalar Wave Solutions

**Ioannis M. Besieris** [1] and **Peeter Saari** [2,3]

[1] The Bradley Department of Electrical and Computer Engineering, Virginia Polytechnic Institute and State University, Blacksburg VA 24060, USA; besieris@gmail.com

[2] Institute of Physics, University of Tartu, W. Ostwaldi 1, 50411, Tartu, Estonia;

[3] Estonian Academy of Sciences, Kohtu 6, 10130, Tallinn, Estonia

**Abstract**

Significant amount of information is available in wave optics about monochromatic paraxial beams characterized by nodal knots, the latter being threads of complete darkness (lines of zero intensity) tied into knots. Also, well -known are finite-energy spatiotemporally localized vector-valued wave solutions characterized by knotted topologies. Unfortunately, except for solutions to the Schrödinger-like equation describing anomalous dispersion, no scalar-valued spatiotemporally localized waves with analogous knotted topological structures are known to date. It is the goal in this article to fill in this gap by deriving finite-energy spatiotemporally localized scalar-valued wave solutions characterized specifically by Hopf and trefoil nodal knot topologies. This will extend to spatiotemporally localized pulsed fields the diverse important applications of monochromatic light beams, having nontrivial topological structures, in trapping and guiding particles, image processing and optical communications; also, in applications to biology and medicine.

**Keywords:** topological structures; Hopf knot; trefoil; Riemann-Silberstein vector; Bateman functions; focus wave modes; X waves; splash mode; wave equation; Schrödinger equation; Klein-Cordon equation

---

## 1. Introduction

*Luminal*, or focus wave modes (FWM), *superluminal*, or X waves (XWs), and *subluminal* spatiotemporally localized solutions to various hyperbolic equations governing acoustic, electromagnetic and quantum wave phenomena have been studied intensively during the past thirty years (see [1-15] for pertinent literature). Further details can be found in two edited monographs on the subject [16,17] and in the recent thorough review article [18]. In general, both linear and nonlinear localized wave (LW) pulses exhibit distinct advantages in comparison to conventional quasi-monochromatic and pulsed signals. Their spatiotemporal confinement and extended field depths render them especially useful in diverse physical applications. Experimental demonstrations have been per-formed in the acoustic and optical regimes [5, 7, 10, 20-23]. Work, however, has been carried out at microwave frequencies recently [24-26].

A class of finite-energy space-time localized electromagnetic waves is known to have interesting topological properties, specifically with respect to linkages of electric and magnetic field lines. Among them are the basic Hopfion, as well as higher-order Hopf-Ranada solutions [27-31].

An outline of our knowledge of finite-energy spatiotemporally localized knotted *vector-valued* wave solutions is given in Section 2. Unfortunately, although huge amount of information is known in wave optics about monochromatic paraxial beams characterized by nodal knots – the latter being threads of complete darkness (lines of zero intensity) tied into knots [32-44], no scalar-valued spatiotemporally localized waves with analogous knotted topological structures are known to date. It is the goal in this article to fill in this gap by deriving finite-energy spatiotemporally localized scalar-valued wave solutions characterized especially by Hopf and trefoil nodal knot topologies. This work is presented and discussed in Sections 3-8. Concluding remarks are made in Section 9.

## 2. Finite-energy spatiotemporally localized knotted vector-valued wave solutions

*2.1 Bateman Conjugate Functions*

Possibly motivated by Whittaker's two-scalar potential theory [45], Bateman [46] introduced two complex-valued scalar functions $\alpha\left(\vec{r},t\right)$ and $\beta\left(\vec{r},t\right)$ that can be used to derive electromagnetic fields in vacuum characterized by interesting topological properties. Known as the *Bateman conjugate functions,* an arbitrary functional $\phi\left[\alpha\left(\vec{r},t\right),\beta\left(\vec{r},t\right)\right]$ of them obeys the nonlinear characteristic (eikonal) equation

$$\left(\frac{\partial\phi}{\partial x}\right)^2+\left(\frac{\partial\phi}{\partial y}\right)^2+\left(\frac{\partial\phi}{\partial z}\right)^2-\frac{1}{c^2}\left(\frac{\partial\phi}{\partial t}\right)^2=0. \tag{1}$$

Furthermore, two arbitrary functionals $\phi_{1,2}\left[\alpha\left(\vec{r},t\right),\beta\left(\vec{r},t\right)\right]$ obey the *Bateman constraint relation*

$$\nabla\phi_1\times\nabla\phi_2=i\frac{1}{c}\left(\frac{\partial\phi_1}{\partial t}\nabla\phi_2-\frac{\partial\phi_2}{\partial t}\nabla\phi_1\right). \tag{2}$$

*2.2 Riemann-Silberstein Complex Vector*

*Real-valued* electromagnetic fields $\vec{E}\left(\vec{r},t\right)$ and $\vec{B}\left(\vec{r},t\right)$ governed by Maxwell equations in vacuum are conveniently expressed in terms of a complex-valued vector

$$\vec{F}=\sqrt{\frac{\varepsilon_0}{2}}\left(\vec{E}+ic\vec{B}\right). \tag{3}$$

The *Riemann-Silberstein vector* [47,48] obeys the equations

$$\nabla\times\vec{F}=i\frac{1}{c}\frac{\partial}{\partial t}\vec{F},\quad \nabla\cdot\vec{F}=0, \tag{4}$$

which are exactly equivalent to the original Maxwell equations for the real fields.

Several important physical quantities associated with the real fields can be expressed conveniently in terms of $\vec{F}\left(\vec{r},t\right)$. Specifically, the Poynting vector, the electromagnetic field energy density, the electromagnetic momentum density and the electromagnetic angular momentum density can be written as

$$
\begin{aligned}
\vec{P} &= \vec{E}\times\vec{H} = -ic\vec{F}^{*}\times\vec{F},\\
w_{em} &= \frac{1}{2}\left(\varepsilon_0\vec{E}\cdot\vec{E}+\mu_0\vec{H}\cdot\vec{H}\right)=\vec{F}^{*}\cdot\vec{F},\\
\vec{M} &= \frac{1}{c^2}\vec{E}\times\vec{H} = -i\frac{1}{c}\vec{F}^{*}\times\vec{F},\\
\vec{J} &= \vec{r}\times\vec{M} = -i\frac{1}{c}\vec{r}\times\left(\vec{F}^{*}\times\vec{F}\right).
\end{aligned}
\tag{5}
$$

For the following discussion, the Riemann-Silberstein vector is related to the Bateman conjugate functions in terms of a complex-valued vector potential as follows:

$$
\begin{aligned}
\vec{F} &= \sqrt{\frac{\varepsilon_0}{2}}c\nabla\times\vec{A} = \frac{1}{2}\sqrt{\varepsilon_0}c\alpha^{p-1}\beta^{q-1}\nabla\alpha\times\nabla\beta;\\
\vec{A}_{pq} &= \left(\varepsilon_0/2\right)^{1/2}c\left(pq\right)^{-1}\alpha^{p}\nabla\beta^{q} = \vec{A}_e + i\vec{A}_m;\ p,q=1,2,3,\ldots
\end{aligned}
\tag{6}
$$

By virtue of its construction, the Riemann-Silberstein vector in Equation (6) obeys Equations (4), and from its definition in Equation (3) it follows that $\vec{E}=c\nabla\times\vec{A}_e,\ \vec{B}=\nabla\times\vec{A}_m$. Another consequence of the specific construction of the Riemann-Silberstein vector according to Equation (6) is the *nullity* property: 76

$$
\begin{aligned}
\vec{F}\left(\vec{r},t\right)\cdot\vec{F}\left(\vec{r},t\right) &= i\varepsilon_0 I_2 + \frac{\varepsilon_0}{2}I_1 = 0;\\
I_1 &= \left|\vec{E}\left(\vec{r},t\right)\right|^2 - c^2\left|\vec{B}\left(\vec{r},t\right)\right|^2 = 0,\\
I_2 &= \vec{E}\left(\vec{r},t\right)\cdot\vec{B}\left(\vec{r},t\right) = 0.
\end{aligned}
\tag{7}
$$

It requires, specifically, that the two relativistic invariants $I_{1,2}$ vanish identically. The corresponding real fields $\vec{E}\left(\vec{r},t\right)$ and $\vec{B}\left(\vec{r},t\right)$ are called *null electromagnetic fields*. Such fields are transverse with respect to the local flow of energy, which is equipartitioned between the electric and magnetic fields, and the modulus of the local energy transport velocity, defined as the ratio of the Poynting vector and the electromagnetic energy density, equals the speed of light in vacuo, although this velocity can vary both spatially and temporally. The simplest electromagnetic null field is a plane wave. More complex null fields are characterized by "helical" or vortex structures on planes transverse to the direction of propagation, and, in general, are relatively simple so that explicit calculations can be made of the total energy and the total angular momentum they carry.

### *2.3 Specific Bateman Conjugate Functions*

Irvine and Bouwmeester [27] explored the physical properties of an exceptional luminal localized solution to the homogeneous Maxwell equations *in vacuo* that exhibits linked and knotted field line structures. The Irvine-Bouwmeester solution was constructed originally by Ranãda [28] using Hopf maps (*fibrations*) and belongs to a more general class of solutions to linear and nonlinear equations known as *Hopfions*. Besieris and Shaarawi [29] have linked the Irvine-Bouwmeester expression to a known null luminal electromagnetic solution.

Consider the luminal Bateman conjugate functions

$$\alpha = \frac{x^2 + y^2 + (z + ia)^2 - c^2t^2}{x^2 + y^2 + z^2 - (ct - ia)^2},$$
$$\beta = \frac{2(x + iy)}{x^2 + y^2 + z^2 - (ct - ia)^2}, \tag{8}$$

where $a$ is a constant parameter. It should be noted that they qualify as Euler potentials because the vector $\nabla\alpha \times \nabla\beta$ has zero divergence. Any functionals of these two functions obey the nonlinear characteristic equation and the Bateman constraint. With $c = 1, a = 1$ (See Ref. [30]; also [31]), they have the property $\alpha\alpha^* + \beta\beta^* = 1$. The surface $|\alpha| =$ *const*. represents a torus for any value of $t$. The surface $|\alpha^2 - \beta^2| =$ *const.* represents two linked tori, $|\alpha^2 - \beta^3| =$ *const*. represents a trefoil knot, and $|\alpha^2 - \beta^5| =$ *const.* depicts a cinquefoil, for any value of time (see Figure 1).

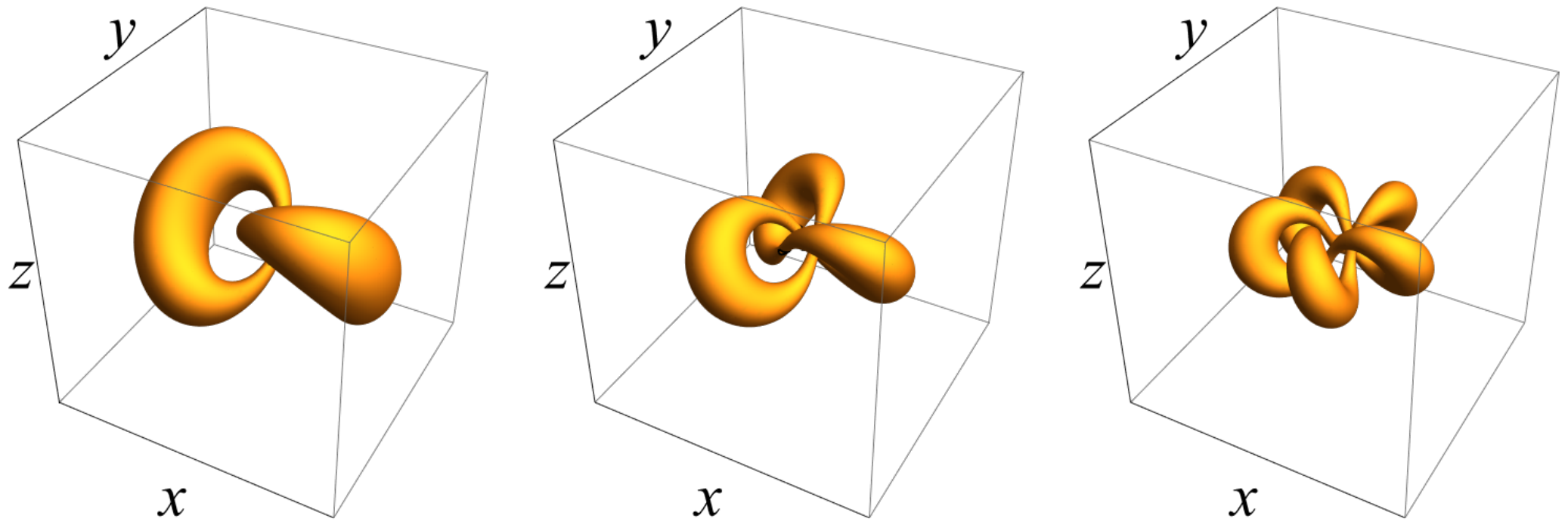

**Figure 1.** Plots of $|\alpha^p - \beta^q| = const.$ for $p = q = 2$ (torus link), $p = 2, q = 3$ (trefoil) and $p = 2, q = 5$ (cinquefoil) at $t = 0$.

When the expressions for the Bateman conjugate functions given in Equation (8) are used in Equation (6), the case $p = 1$ and $q = 1$, corresponding to simplest Hopf-Ranãda solution, or the basic Hopfion, yields linked electromagnetic fields. Any two electric field lines are linked, any two magnetic field lines are linked, and any electric field line is linked to any magnetic field line for any instant of time. Electric and magnetic field line linkages for $p = 1, q = 1$ and $p = 2, q = 3$ are shown in Figure 2. The corresponding fields have finite energy.

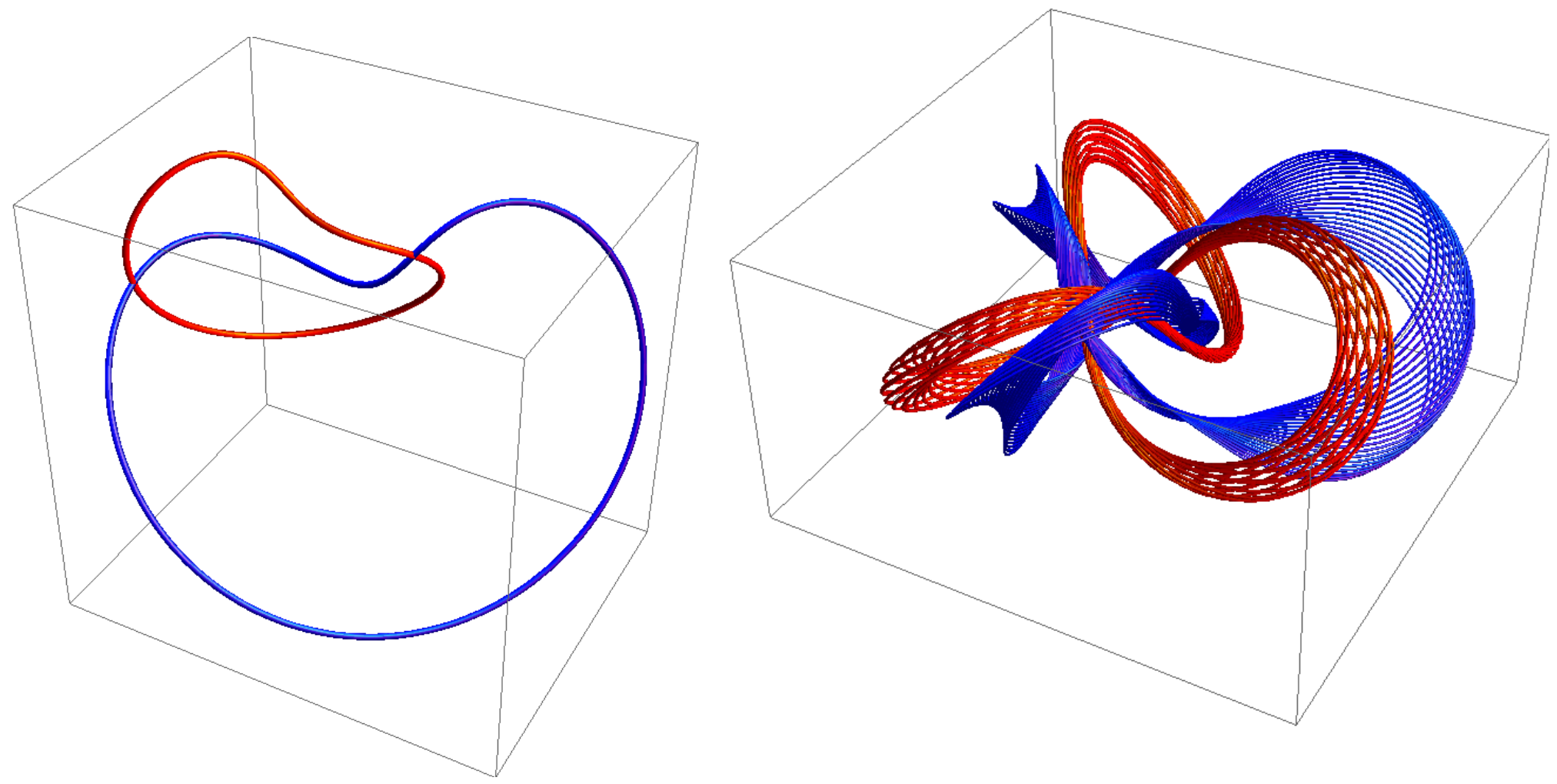

**Figure 2**. Plots of $\vec{E}$ (red) and $\vec{B}$ (blue) field line linkages for $p = q = 1$ (left) and $p = 2, q = 3$ (right) at $t = 0$.

## 3. Hopf nodal knot for a spatiotemporally localized splash mode

### *3.1 Time-independent Bateman conjugate functions*

At $t = 0$, the expressions $\alpha$ and $\beta$ in Equation (8), with $c = 1, a = 1$, are denoted by $u$ and $v$. The latter were used by Brauer in 1928 [49] to construct explicit realizations of complex-valued scalar knotted fields for each (*p, q*) torus knot or link $\left|u^p - v^q\right|$ = *const*. The surface $\left|u^2 - v^2\right|$ = *const*. represents two linked tori and $\left|u^2 - v^3\right|$ = *const*. represents a trefoil linkage.

### *3.2 Hopf nodal knot for the paraxial equation*

Consider the expression

$$u^2 - v^2 = \frac{N(x,y,z)}{D(x,y,z)} = \frac{-4(x+iy)^2 + \left(x^2+y^2+(i+z)^2\right)^2}{\left(1+x^2+y^2+z^2\right)^2}. \tag{9}$$

The numerator contains the information on the Hopf nodal linkage as shown in the figure below.

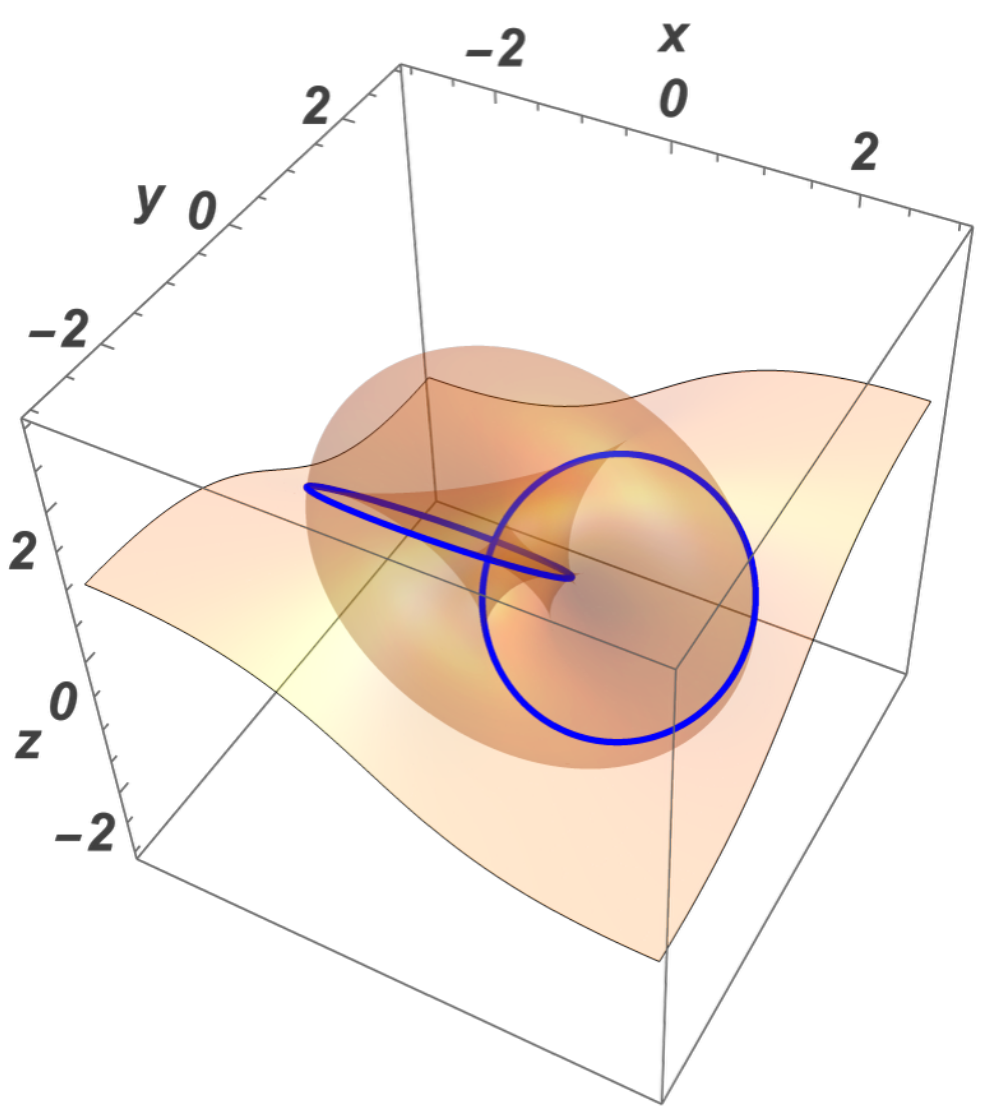

**Figure 3.** Hopf nodal linkage formed from the intersection of the real and imaginary parts of the numerator $N(x,y,z)$; specifically, $\mathrm{Re}\{N(x,y,z)\}=0$ and $\mathrm{Im}\{N(x,y,z)\}=0$.

Next, a paraxial nodal Hopf knot is formed as follows. First the Fourier transform

$$\hat{\psi}(k_x,k_y)=F\left\{N(x,y,0)\exp\left[-\kappa\left(x^2+y^2\right)\right]\right\} \tag{10}$$

is computed (see [38, 39] for an alternative approach). Then, a specific paraxial Hopf nodal knot is obtained by means of the Fourier inversion

$$\begin{aligned}\psi(x,y,z)&=F^{-1}\left\{\hat{\psi}(k_x,k_y)\exp\left[-i\frac{z}{2}\left(k_x^2+k_y^2\right)\right]\right\}\\&=\frac{1}{(1+i2\kappa z)^5}\exp\left[-\kappa\frac{x^2+y^2}{1+i2\kappa z}\right]\\&\times\begin{pmatrix}x^4+y^4+8ixy(i-2z\kappa)^2+y^2\left(2+8iz(i+\kappa)-8z^2\kappa(2+\kappa)\right)\\+(i-2z\kappa)^2\left(-1-4iz(-1+\kappa)+4z^2\left(2+(-2+\kappa)\kappa\right)\right)\\+2x^2\left(y^2+(-i+2z\kappa)\left(-3i+z(-4+6\kappa)\right)\right)\end{pmatrix}.\end{aligned} \tag{11}$$

It obeys the paraxial equation

$$2i\frac{\partial}{\partial z}\psi(x,y,z)+\nabla_t^2\psi(x,y,z)=0. \tag{12}$$

The equality

$$N(x,y,z)=-4(x+iy)^2+\left(x^2+y^2+(i+z)^2\right)^2 \tag{13}$$

in Equation (9) is known as the Hopf Milnor polynomial. It does not satisfy the paraxial equation. However, the procedure followed above allows one to obtain an extended polynomial that satisfies the paraxial equation and, also, inherits the nodal topology of the Hopf Milnor polynomial. If $\kappa$ is set equal to zero in Equation (11), one obtains

$$w(x,y,z)=1+x^4-8ixy+y^4+2x^2\left(y^2-i(-3i-4z)\right)\\ +y^2(2+8iz)-4iz-8z^2. \tag{14}$$

This expression, known as the Hopf paraxial polynomial, obeys the paraxial equation (12). (see [35], [38] and [39] for alternative derivations of such polynomials). Of course, for $z=0$, it is equal to $N(x,y,0)$. However, $N(x,y,z)$ and $w(x,y,z)$ differ significantly in their dependence on $z$. The former is a polynomial of order four in $z$, whereas the latter is of order two. Once the Hopf paraxial polynomial is known, a paraxial Hopf nodal knot -- different from the one in Equation (11) but having the same topology -- can be derived by a different approach based on a conformal symmetry of the Gaussian paraxial solution. Specifically,

$$\phi(x,y,z)=\frac{1}{a_1+iz}\exp\left(-\frac{1}{2}\frac{x^2+y^2}{a_1+iz}\right)w\left(\frac{a_1x}{a_1+iz},\frac{a_1y}{a_1+iz},\frac{a_1z}{a_1+iz}\right), \tag{15}$$

where $a_1$ is a positive free parameter.

*3.3 Hopf nodal knot for the focus wave solution of the scalar wave equation*

In the Hopf Milnor polynomial in Equation (15), $z$ is replaced by $\varsigma/2b$, where $\varsigma=z-t$ and $b$ is a free positive parameter. Using the change $w(x,y,z)\rightarrow W(x,y,\varsigma)$, the expression

$$f(x,y,\varsigma,\eta)=\frac{a_1}{a_1+i\varsigma}exp(ib\eta)\exp\left(-b\frac{x^2+y^2}{a_1+i\varsigma}\right)\\ \times W\left(\frac{a_1x}{a_1+i\varsigma},\frac{a_1y}{a_1+i\varsigma},\frac{a_1\varsigma}{a_1+i\varsigma}\right), \tag{16}$$

with $\eta=z+t$, is a solution of the scalar wave equation, viz.,

$$\left(\nabla_t^2+4\frac{\partial^2}{\partial\varsigma\partial\eta}\right)f(x,y,\varsigma,\eta)=0, \tag{17}$$

that retains the Hopf nodal knot topological structure with respect to the variables $x,y$ and $\varsigma$.

*3.4 Hopf nodal knot for the fine-energy splash mode solution of the scalar wave equation*

Although the solution in Equation (16) is spatiotemporally localized, it has infinite energy. A finite-energy solution can be derived by an appropriate superposition over the free parameter $b$. A particularly simple exact solution can be derived by multiplying the right-hand side of Equation (16) by $b^2\exp(-a_2b)$, with $a_2>0$, and carrying out the integrations

$$I_1 = \frac{a_1}{a_1 + i\varsigma}\int_0^\infty db\, b^2 \exp\left[-b\left(a_2 - i\eta + \frac{x^2+y^2}{a_1+i\varsigma}\right)\right]\left(-\frac{2a_1^2\varsigma^2}{b^2(a_1+i\varsigma)}\right);$$

$$I_2 = \frac{a_1}{a_1 + i\varsigma}\int_0^\infty db\, b^2 \exp\left[-b\left(a_2 - i\eta + \frac{x^2+y^2}{a_1+i\varsigma}\right)\right] \times\left(\frac{-2ia_1\varsigma + 4ia_1\left(\frac{a_1^2(x^2+y^2)}{(a_1+i\varsigma)^2}\right)\varsigma}{b(a_1+i\varsigma)}\right); \tag{18}$$

$$I_3 = \frac{a_1}{a_1 + i\varsigma}\int_0^\infty db\, b^2 \exp\left[-b\left(a_2 - i\eta + \frac{x^2+y^2}{a_1+i\varsigma}\right)\right] \times\left(1 - 2\frac{a_1^2(x^2+y^2)}{(a_1+i\varsigma)^2} + \left(\frac{a_1^2(x^2+y^2)}{(a_1+i\varsigma)^2}\right)^2 - 4\left(\frac{a_1(x+iy)}{a_1+i\varsigma}\right)^2\right).$$

The three integrals can be carried out explicitly using a table of Laplace transforms, resulting in the Hopf nodal splash mode solution

$$I_H(x,y,\varsigma,\eta) = I_1 + I_2 + I_3. \tag{19}$$

Setting $\eta = \varsigma + 2t$, the intersection of the real and imaginary parts of $I_H(x,y,\varsigma,t)$ is shown below relative to the pulse center $(\varsigma = 0)$ for two values of $t$.

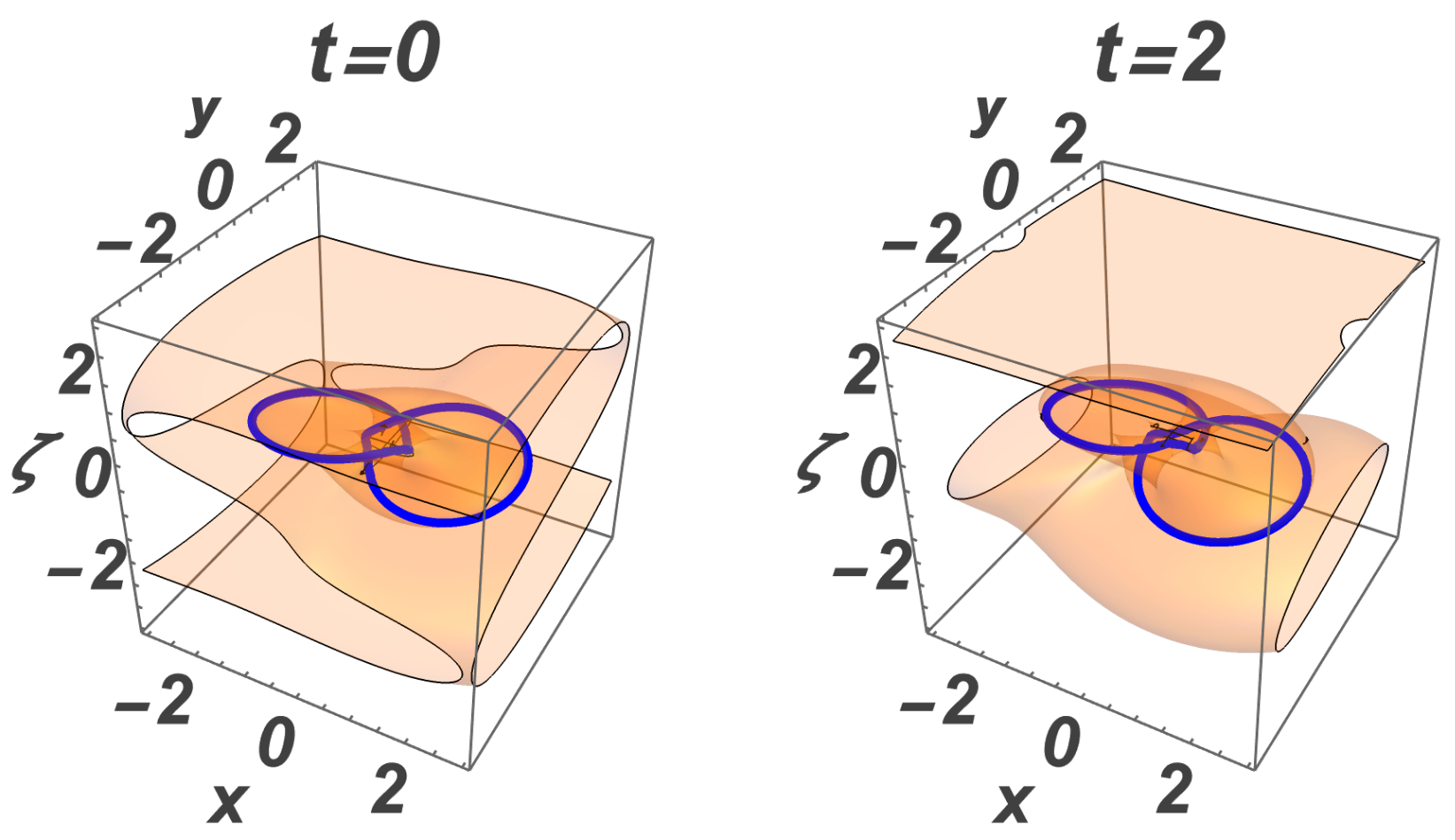

**Figure 4.** Hopf nodal linkage formed from the intersection of the real and imaginary parts of the scalar splash mode Hopf nodal knot solution for $t = 0$ and $t = 2$, with parameter values $a_1 = 1$ and $a_2 = 10$.

Due to its bidirectionality, the spatiotemporal localization, as well as the Hopf nodal knot topology, of the solution $I_H(a_1, a_2, x, y, \varsigma; t)$ cannot be maintained for relatively large values of $t$. However, for $a_2 >> a_1$, the pulse is esentially invariant.

The energy density of the scalar splash mode Hopf nodal knot solution is given by

$$U_H(x, y, \varsigma, \eta) = \frac{1}{2}\left(\frac{\partial}{\partial t} I_H \frac{\partial}{\partial t} I_H^* + \nabla I_H \cdot \nabla I_H^*\right). \tag{20}$$

Resorting to $\eta = \varsigma + 2t$, plots of $U_H(x, y, \varsigma, t)$ are shown in the figure below for three values of time.

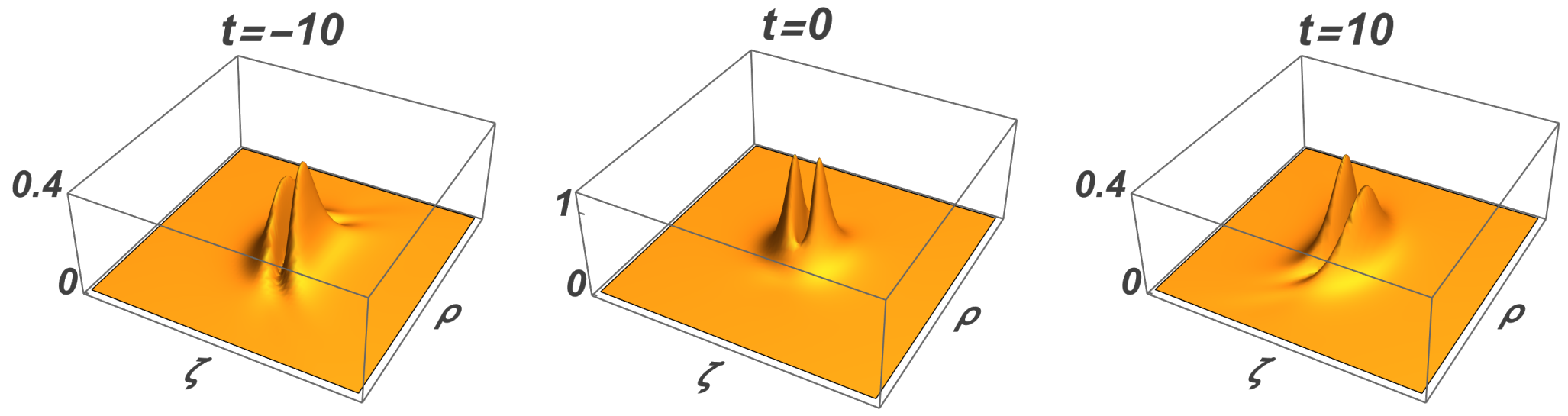

**Figure 5.** Energy density of the scalar splash mode Hopf nodal knot solution for $t = -10, 0, 10$, with parameter values $a_1 = 1$ and $a_2 = 10$.

## 4. Hopf nodal knot for a finite-energy splash mode solution of the scalar wave equation based on a superposition of Gauss-Laguerre focus wave modes

The numerator in Equation (9) at $z = 0$ is expressed in polar form, viz.,

$$N(\rho, \phi, 0) = 1 - 2\rho^2 - 4\rho^2 e^{2i\phi} + \rho^4, \tag{21}$$

and multiplied by a Gaussian function as follows:

$$F_H(\rho, \phi; w) = N(\rho, \phi, 0) \exp\left[-\rho^2 / (2w)\right], \tag{22}$$

where $w$ is a constant. Using a standard procedure (see, e. g., [35, 36, 43]), $F_H(\rho, \phi)$ is used in a superposition of Gauss-Laguerre modes as

$$\begin{aligned} f_H(\rho, \phi, z) &= \sum_{m,n} c_{mn} LG_{m,n}(\rho, \phi, z; w_0); \\ c_{mn} &= \iint d^2 r_t LG_{-m,n}(\rho, \phi, 0; w) LG_{0,0}(\rho, \phi, 0; w) F_H(\rho, \phi; w), \end{aligned} \tag{23}$$

with the Gauss-Laguerre modes defined by

$$\begin{aligned} LG_{m,n}(\rho, \phi, z; w_0) &= \sqrt{\frac{n!}{\pi(|m| + n)!}} \frac{\rho^{|m|} e^{im\phi}}{w_0^{|m|+1}} \frac{(1 - iz/z_R)^n}{(1 + iz/z_R)^{n+|m|+1}} \\ &\times \exp\left[-\frac{\rho^2}{2w_0^2(1 + iz/z_R)}\right] L_n^{|m|}\left(\frac{\rho^2}{w_0^2(1 + z^2/z_R^2)}\right); \; z_R = k_0 w_0^2. \end{aligned} \tag{24}$$

The coefficients required in Equation (23) are given as follows:

$$c_{00}=1-2w^2+2w^4,\ c_{01}=2\left(w^2-2w^4\right),\ c_{02}=2w^4,\ c_{20}=-4\sqrt{2}w^2. \tag{25}$$

With the changes $k=1,\ w_0=\sqrt{a_1/(2b)},\ z\to\varsigma/\left(2b\right)$, where $a_1$ and $b$ are free parameters, the expression

$$\begin{aligned} g_H\left(\rho,\phi,\varsigma,\eta\right) &= e^{ib\eta}f_H\left(\rho,\phi,\varsigma\right) \\ &= \sqrt{\frac{2}{\pi}}\left(\begin{aligned} &\frac{\sqrt{a_1}}{\left(a_1+i\varsigma\right)^3}(a_1^2-2ia_1\left(-1+2w_2\right)\varsigma+\left(-1+4w^2-8w^4\right)\varsigma^2 \\ &-\frac{4a^{3/2}}{\left(a_1+i\varsigma\right)^4}bw^2\left(a_1+2a_1e^{2i\phi}+i\left(1+2e^{2i\phi}-4w^2\right)\varsigma\right)\rho^2 \\ &+\frac{4a^{5/2}}{\left(a_1+i\varsigma\right)^5}w^4\rho^4 \end{aligned}\right) \\ &\times\exp\left[ib\eta-b\frac{\rho^2}{a_1+i\varsigma}\right] \end{aligned} \tag{26}$$

is an infinite-energy superposition of focus wave mode solutions to the scalar wave equation [see Equation (17)] characterized by a Hopf nodal link topology. A finite-energy spatiotemporally localized splash mode maintaining the same topology can be found by a superposition over the free parameter $b$ using the spectrum $\exp\left(-a_2b\right)$, with $a_2>0$. Specifically,

$$\begin{aligned} h_H\left(\rho,\phi,\varsigma,\eta\right) &= \int_0^\infty db\, e^{-a_2b}g_H\left(\rho,\phi,\varsigma,\eta\right) \\ &= \sqrt{\frac{2}{\pi}}\sqrt{a_1}\frac{1}{\left(a_1+i\varsigma\right)^3\left(a_2-i\eta+\dfrac{\rho^2}{a_1+i\varsigma}\right)} \\ &\times\left(a_1^2-2ia_1\left(1+2w^2\right)\varsigma+\left(-1+4w^2-8w^4\right)\varsigma^2\right) \\ &+\sqrt{\frac{2}{\pi}}4a_1^{3/2}\frac{1}{\left(a_1+i\varsigma\right)^4\left(a_2-i\eta+\dfrac{\rho^2}{a_1+i\varsigma}\right)^2} \\ &\times w^2\left(a_1+2a_1e^{2i\phi}+i\left(1+2e^{2i\phi}-4w^2\right)\varsigma\right)\rho^2 \\ &+\sqrt{\frac{2}{\pi}}4a_1^{5/2}\frac{w^4\rho^4}{\left(a_1+i\varsigma\right)^5\left(a_2-i\eta+\dfrac{\rho^2}{a_1+i\varsigma}\right)^3}. \end{aligned} \tag{27}$$

Setting $\eta=\varsigma+2t$, the intersection of the real and imaginary parts of $h_H\left(x,y,\varsigma,t\right)$ is shown below for two values of $t$.

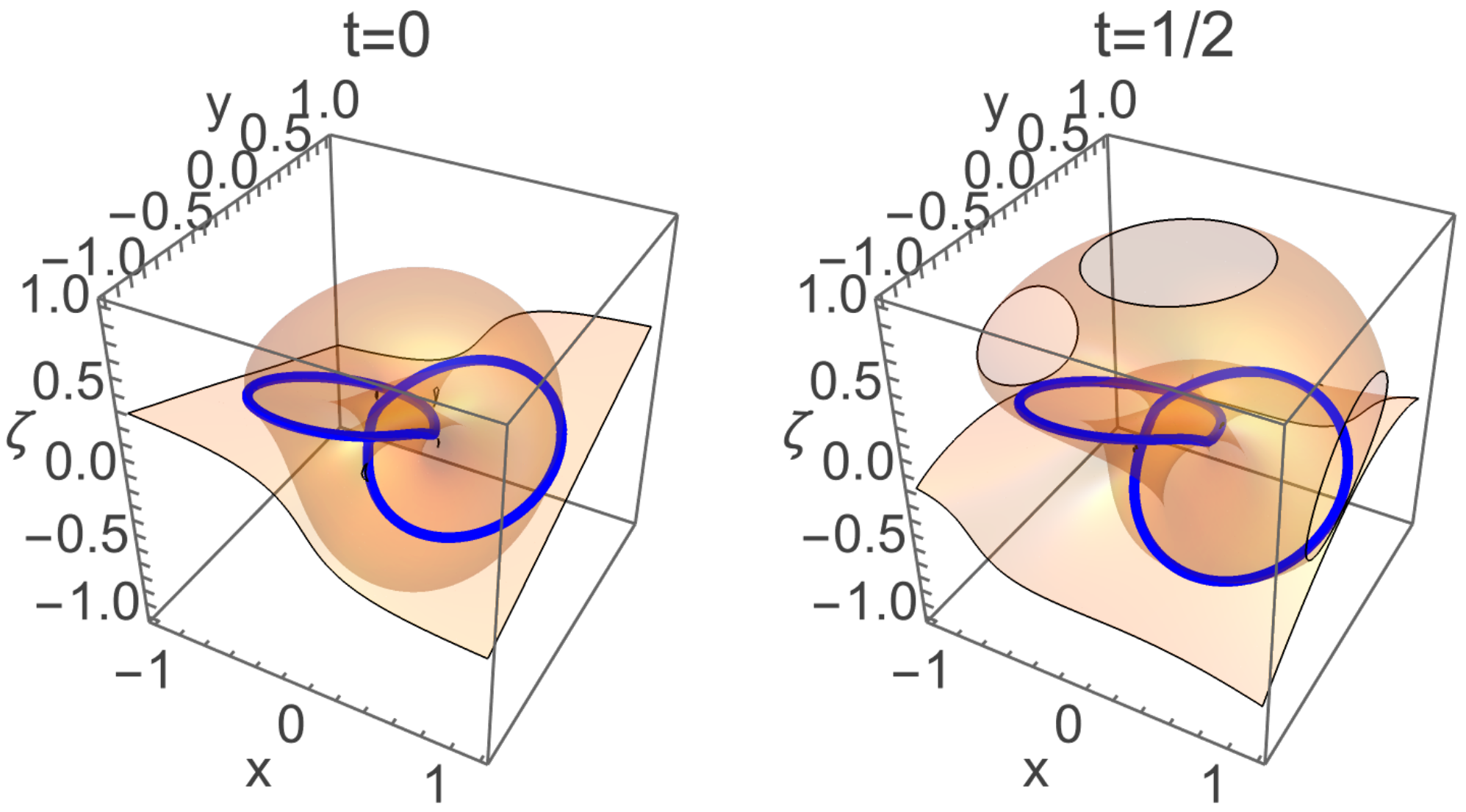

**Figure 6.** Hopf nodal linkage formed from the intersection of the real and imaginary parts of the scalar splash mode Hopf nodal knot solution for $t=0$ and $t=1/2,$ with parameter values $a_1=1,\ a_2=2$ and $w=2.$

## 5. Trefoil nodal knot for a spatiotemporally localized splash mode

Consider the expression

$$u^2-v^3=\frac{N(x,y,z)}{D(x,y,z)}$$
$$=\frac{-8(x+iy)^3+\left(1+x^2+y^2+z^2\right)\left(x^2+y^2+(i+z)^2\right)^2}{\left(1+x^2+y^2+z^2\right)^3}. \tag{28}$$

The numerator contains the information on the trefoil nodal linkage as shown in the figure below

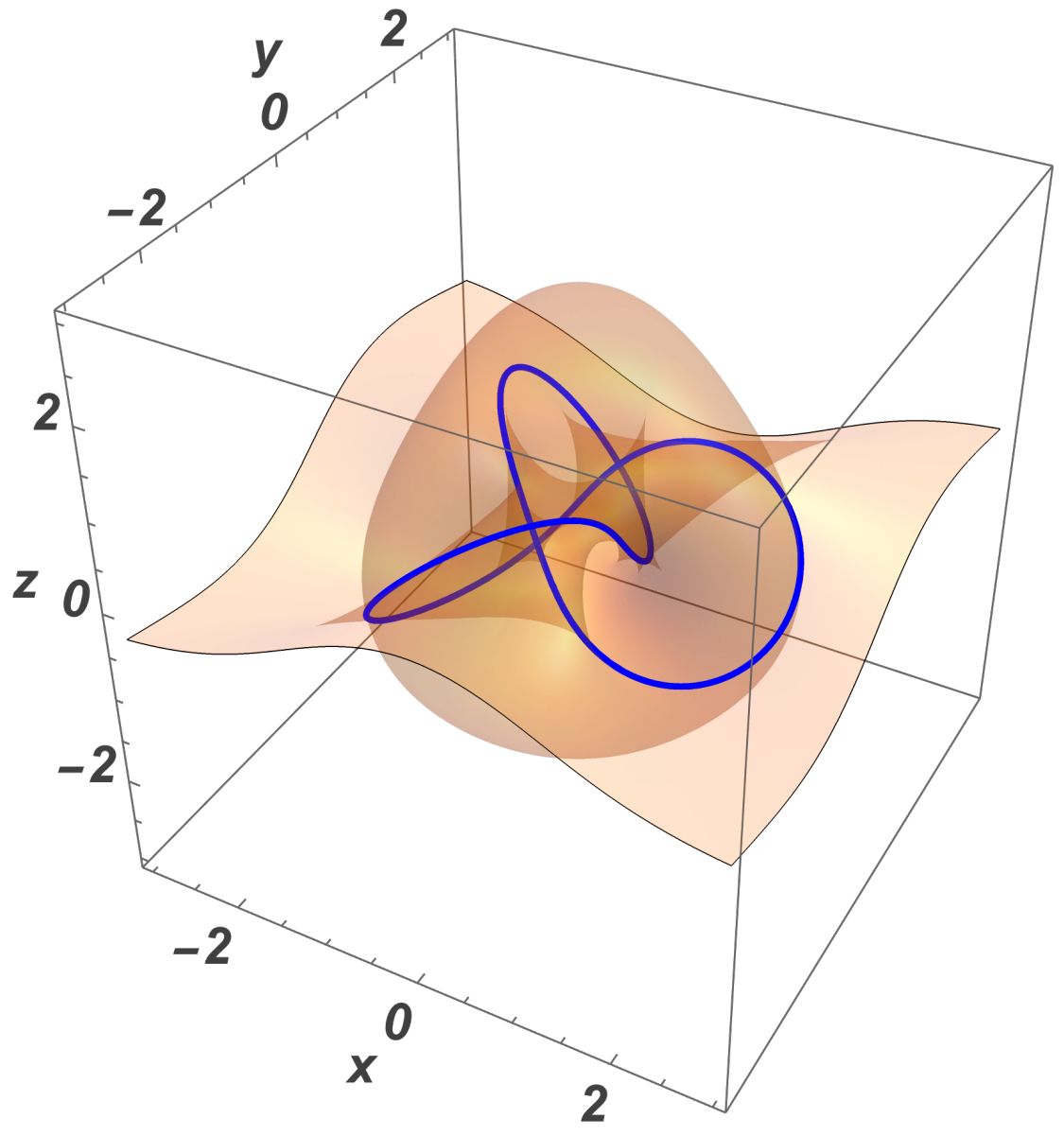

**Figure 7.** Trefoil nodal linkage formed from the intersection of the real and imaginary parts of the numerator $N(x,y,z)$.

The paraxial trefoil polynomial can be derived by a procedure analogous to that followed in the previous section (see, also, [35]). It is given explicitly as follows:

$$\begin{aligned} w(x,y,z) &= 1-8x^3+x^6+24xy^2+8iy^3 \\ &+y^6+y^4(-1+18iz)+x^4(-1+3y\wedge 2+18iz) \\ &-2iz+8z^2-48iz^3+x^2\begin{pmatrix} -1-24iy+3y^4 \\ +y^2(-2+36iz)-8iz-72z^2 \end{pmatrix} \\ &-y^2(1+8iz+72z^2). \end{aligned} \quad (29)$$

This is converted to $W(x,y,\varsigma)$ and the expression $f(x,y,\varsigma,\eta)$ in Equation (16) satisfies the scalar wave equation for a normalized speed $c=1$ in vacuum.

An exact finite-energy splash mode solution exhibiting a trefoil nodal knot topology can be derived by multiplying the right-hand side of Equation (16) by $b^3\exp(-a_2b)$, with $a_2>0$, and carrying out the integrations

$$I_1 = \frac{a_1}{a_1 + i\varsigma} \int_0^\infty db\, b^3 \exp\left[ -b\left( a_2 - i\eta + \frac{x^2 + y^2}{a_1 + i\varsigma} \right)\right]\left( -\frac{6ia_1^3\varsigma^3}{b^3\left(a_1 + i\varsigma\right)^3} \right);$$

$$I_2 = \frac{a_1}{a_1 + i\varsigma} \int_0^\infty db\, b^3 \exp\left[ -b\left( a_2 - i\eta + \frac{x^2 + y^2}{a_1 + i\varsigma} \right)\right] \times\left( \frac{2a_1^2\varsigma^2 - 18a_1^4\left( \frac{x^2 + y^2}{\left(a_1 + i\varsigma\right)^2} \right)\varsigma^2}{b^2\left(a_1 + i\varsigma\right)^2} \right);$$

$$I_3 = \frac{a_1}{a_1 + i\varsigma} \int_0^\infty db\, b^3 \exp\left[ -b\left( a_2 - i\eta + \frac{x^2 + y^2}{a_1 + i\varsigma} \right)\right] \times\left( \frac{-ia_1\varsigma + 4ia_1\left( \frac{a_1^2\left(x^2 + y^2\right)}{\left(a_1 + i\varsigma\right)^2} \right)\varsigma - 9ia_1\left( \frac{a_1^2\left(x^2 + y^2\right)}{\left(a_1 + i\varsigma\right)^2} \right)^2\varsigma}{b\left(a_1 + i\varsigma\right)} \right); \tag{30}$$

$$I_4 = \frac{a_1}{a_1 + i\varsigma} \int_0^\infty db\, b^3 \exp\left[ -b\left( a_2 - i\eta + \frac{x^2 + y^2}{a_1 + i\varsigma} \right)\right] \times\left( \begin{array}{l} 11 - \left( \frac{a_1^2\left(x^2 + y^2\right)}{\left(a_1 + i\varsigma\right)^2} \right)^2 + \left( \frac{a_1^2\left(x^2 + y^2\right)}{\left(a_1 + i\varsigma\right)^2} \right)^3 \\ +8\left( \frac{a_1\left(x + iy\right)}{a_1 + i\varsigma} \right)^3 - \frac{a_1^2\left(x^2 + y^2\right)}{\left(a_1 + i\varsigma\right)^2} \end{array} \right).$$

The four integrals can be carried out explicitly using a table of Laplace transforms, resulting in the scalar splash mode trefoil nodal knot solution

$$I_t\left(x, y, \varsigma, \eta\right) = I_1 + I_2 + I_3 + I_4. \tag{31}$$

Setting $\eta = \varsigma + 2t,$ the intersection of the real and imaginary parts of $I_t\left(x, y, \varsigma; t\right)$ is shown below for two values of $t$.

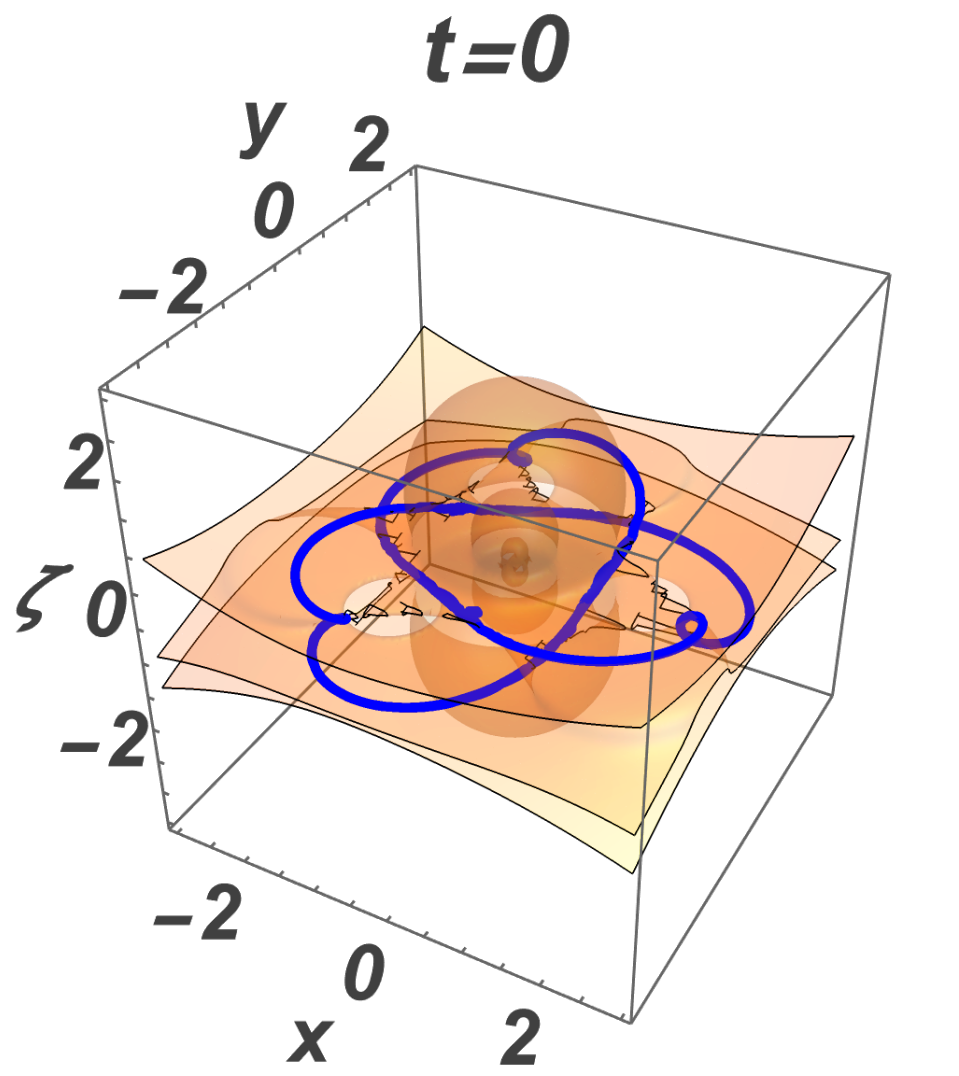

**Figure 8.** Trefoil nodal linkage formed from the intersection of the real and imaginary parts of the scalar splash mode Hopf nodal knot solution for $t=0$ and $t=1$, with parameter values $a_1 = 10^{-1}$ and $a_2 = 1$.

Resorting to $\eta = \varsigma + 2t$, plots of the energy density $U_t\left(x, y, \varsigma, t\right)$ are shown in the figure below for three values of time.

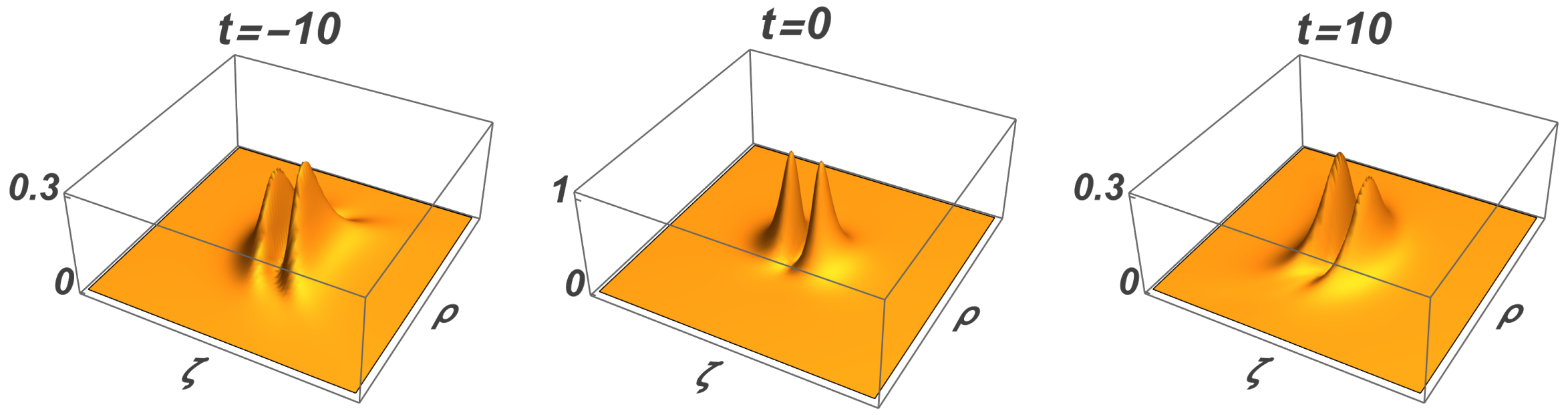

**Figure 9.** Energy density of the scalar splash mode Hopf nodal knot solution for $t=-10, 0, 10$, with parameter values $a_1 = 1$ and $a_2 = 10$.

## 6. Hopf nodal knotted solution of the Schrödinger equation

Consider the time-dependent Bateman conjugate functions in Equation (8) with $a=1$ and $c=1$. Then,

$$\alpha^2 - \beta^2 = \frac{N\left(x, y, z, t\right)}{D\left(x, , z, t\right)} = \frac{-4\left(x+iy\right)^2 + \left(-1-t^2+x^2+y^2+z^2+2iz\right)^2}{\left(-\left(-i+t\right)^2+x^2+y^2+z^2\right)^2}. \qquad (32)$$

The numerator contains the information on the Hopf nodal linkage for all values of time. Figure 3 shows the Hopf nodal knot at $t=0$.

One can proceed to determine a Hopf nodal knotted solution to the Schrödinger equation

$$2i\frac{\partial}{\partial t}\psi\left(\vec{r},t\right)+\nabla^2\psi\left(\vec{r},t\right)=0 \tag{33}$$

by extending the method used in Section 3. Specifically, first the Fourier transform

$$\hat{\psi}\left(k_x,k_y,k_z\right)=F\left\{N\left(x,y,z,0\right)\exp\left[-\kappa\left(x^2+y^2+z^2\right)\right]\right\} \tag{34}$$

is computed. Then, a specific Hopf nodal knot is obtained by means of the Fourier inversion

$$\begin{aligned}\psi(x,y,z,t)&=F^{-1}\left\{\hat{\psi}\left(k_x,k_y,k_z\right)\exp\left[-i\frac{t}{2}\left(k_x^2+k_y^2+k_z^2\right)\right]\right\}\\&=\frac{1}{\left(1+it\right)^{11/2}}\exp\left(-\frac{1}{2}\frac{x^2+y^2+z^2}{1+it}\right)\\&\times\left(\begin{array}{l}6t^4+4t^3\left(-i+4z\right)-t^2\left(-9+12x^2+8ixy+4y^2+28iz+4z^2\right)\\+10it\left(x^2+y^2+\left(i+z\right)^2\right)+\left(x\left(2i+x\right)+\left(2+y\right)y+\left(i+z\right)^2\right)\\\times\left(x\left(-2i+x\right)+y\left(2+y\right)+\left(i+z\right)^2\right)-4t\left(\begin{array}{l}4xy+x^2\left(-i+z\right)+\left(i+z\right)^2\\+y^2\left(3i+z\right)\end{array}\right)\end{array}\right).\end{aligned} \tag{35}$$

The intersection of the real and imaginary parts of this solution is shown in the figure below for two times.

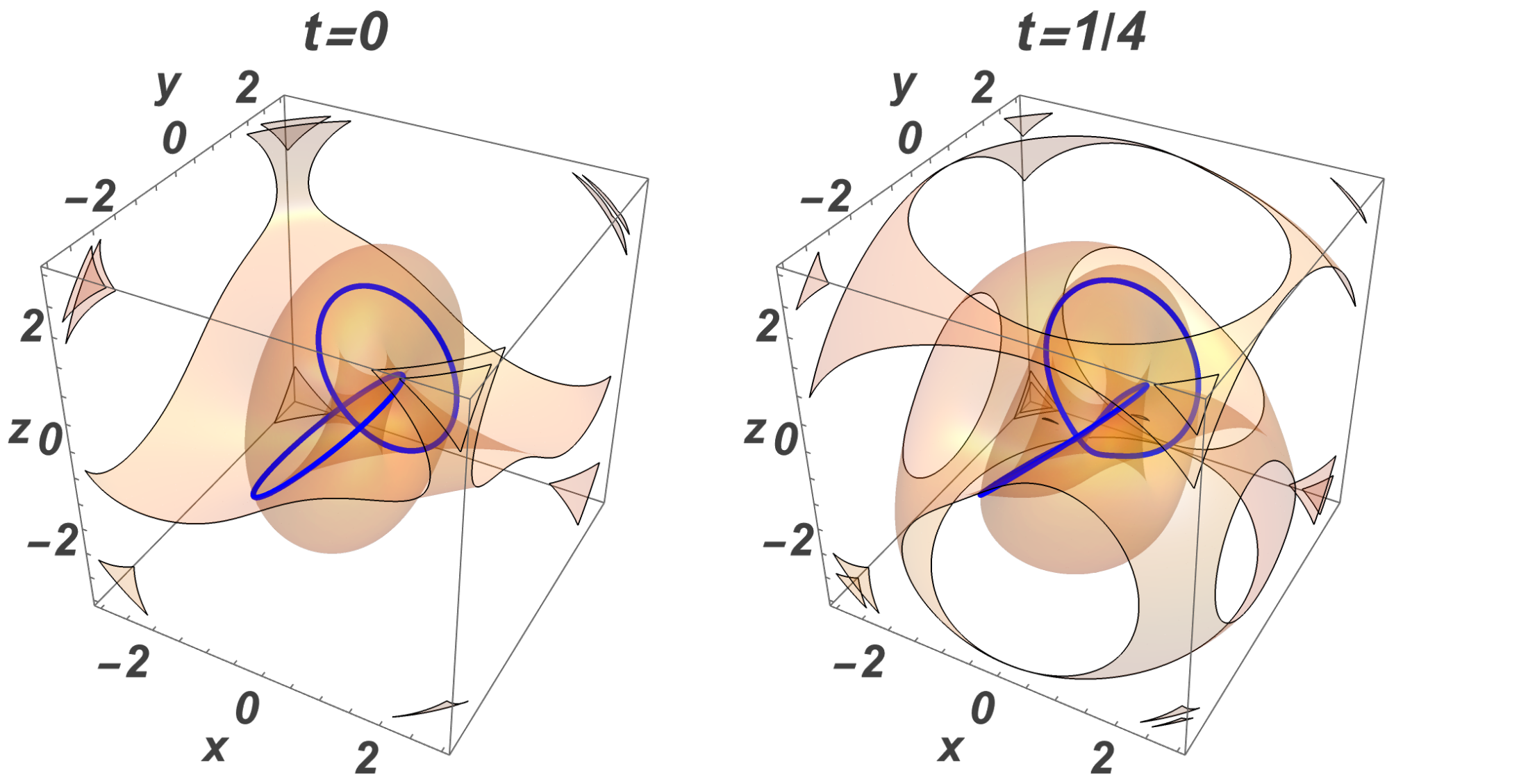

**Figure 10.** The intersection of the real and imaginary parts of the solution in Eq. (29) at $t=0$ and $t=1/4$.

The Schrödinger Hopf polynomial following from the solution in Equation (35) is given by (see, also, [35])

$$w(x,y,z,t) = -1 + 10it + 15t^2 - 2x^2 - 10itx^2 - x^4 - 8ixy$$
$$+ 6y^2 - 10ity^2 - 2x^2y^2 - y^4 + 4iz + 20tz - 4ix^2z \qquad (36)$$
$$- 4iy^2z + 6z^2 - 10itz^2 - 2x^2z^2 - 2y^2z^2 - 4iz^3 - z^4.$$

Of course, for $t = 0$, it is equal to $N(x,y,z,0)$.More complicated solutions, such as a solution characterized by a trefoil nodal knot topology can be derived by extending the procedure followed in Section 3.4.

It should be noted that if the following changes are made in the solution in Equation (35)

$$t \to z,\ z \to Z = \tau / \sqrt{\bar{\beta}_2},\ x \to \sqrt{\beta_0}X, y \to \sqrt{\beta_0}Y; \beta_0 = \beta(\omega_0);$$
$$\tau = t - (z / v_{gr});\ v_{gr} = 1/\beta_1;\ \beta_1 \equiv d\beta(\omega)/d\omega\big|_{\omega=\omega_0} \qquad , \qquad (37)$$
$$\bar{\beta}_2 = |\beta_2|,\ \beta_2 \equiv d^2\beta(\omega)/d\omega^2\big|_{\omega=\omega_0}$$

one would obtain a solution to the anomalous bidispersive equation

$$\left( i\frac{\partial}{\partial z} + \frac{1}{2}\bar{\beta}_2 \frac{\partial^2}{\partial \tau^2} + \frac{1}{2\beta_0}\nabla_\perp^2 \right)\varphi(x,y,\tau,z) = 0 \qquad (38)$$

within the framework of the *slowly varying envelope approximation* (SVEA). In Equation (37), $\beta(\omega)$ is the frequency-dependent wave number, $\omega_0$ is the carrier angular frequency, $v_{gr}$ denotes the group speed and $\beta_2$ is the index of dispersion. The latter is positive for anomalous dispersion and negative for normal dispersion. The solution in Equation (35) is characterized by a Hopf nodal knot topology in the variables $X, Y, Z$ for relatively small values of z (see, e.g., [40-42]).

## 7. Unknot solution of the Klein-Gordon equation

Consider the focus wave mode solution of the scalar wave equation in free space given in Equation (16) of Section 3.3. With the modification shown below, viz.,

$$f(x,y,\varsigma,\eta) = \frac{a_1}{a_1 + i\varsigma} exp(ib\eta)\exp\left[-\frac{p^2}{4b}(a_1 + i\varsigma)\right]\exp\left(-b\frac{x^2+y^2}{a_1+i\varsigma}\right)$$
$$\times W\left(\frac{a_1 x}{a_1+i\varsigma}, \frac{a_1 y}{a_1+i\varsigma}, \frac{a_1\varsigma}{a_1+i\varsigma}\right), \qquad (39)$$

one has a solution of the Klein-Gordon equation

$$\left(\nabla_t^2 + 4\frac{\partial^2}{\partial\varsigma\partial\eta} - p^2\right) f(x,y,\varsigma,\eta) = 0 \qquad (40)$$

governing a cold plasma. It corresponds to a situation where the square of the frequency-dependent wave number (see Section 6) is of second order; specifically, $\beta^2(\omega) = (\omega/c)^2 - (\omega_p/c)^2 = (\omega/c)^2 - p^2;\ p = (\omega_p/c), \omega_p$ being the plasma frequency.

$W(x,y,\varsigma)$ in Equation (39) is an appropriate polynomial satisfying the paraxial equation

$$4i\frac{\partial}{\partial\varsigma}W(x,y,\varsigma)+\nabla_t^2 W(x,y,\varsigma)=0. \tag{41}$$

In the following discussion a finite-energy spatiotemporally localized solution to the Klein-Gordon equation will be determined characterized by a simple unknot topology, essentially a closed loop. Toward this goal, consider the expression

$$u-v=\frac{N(x,y,z)}{D(x,y,z)}=\frac{x^2-2x+2iy+(i+z)}{1+x^2+y^2+z^2}. \tag{42}$$

The numerator contains the information on the unknot property shown in the figure below.

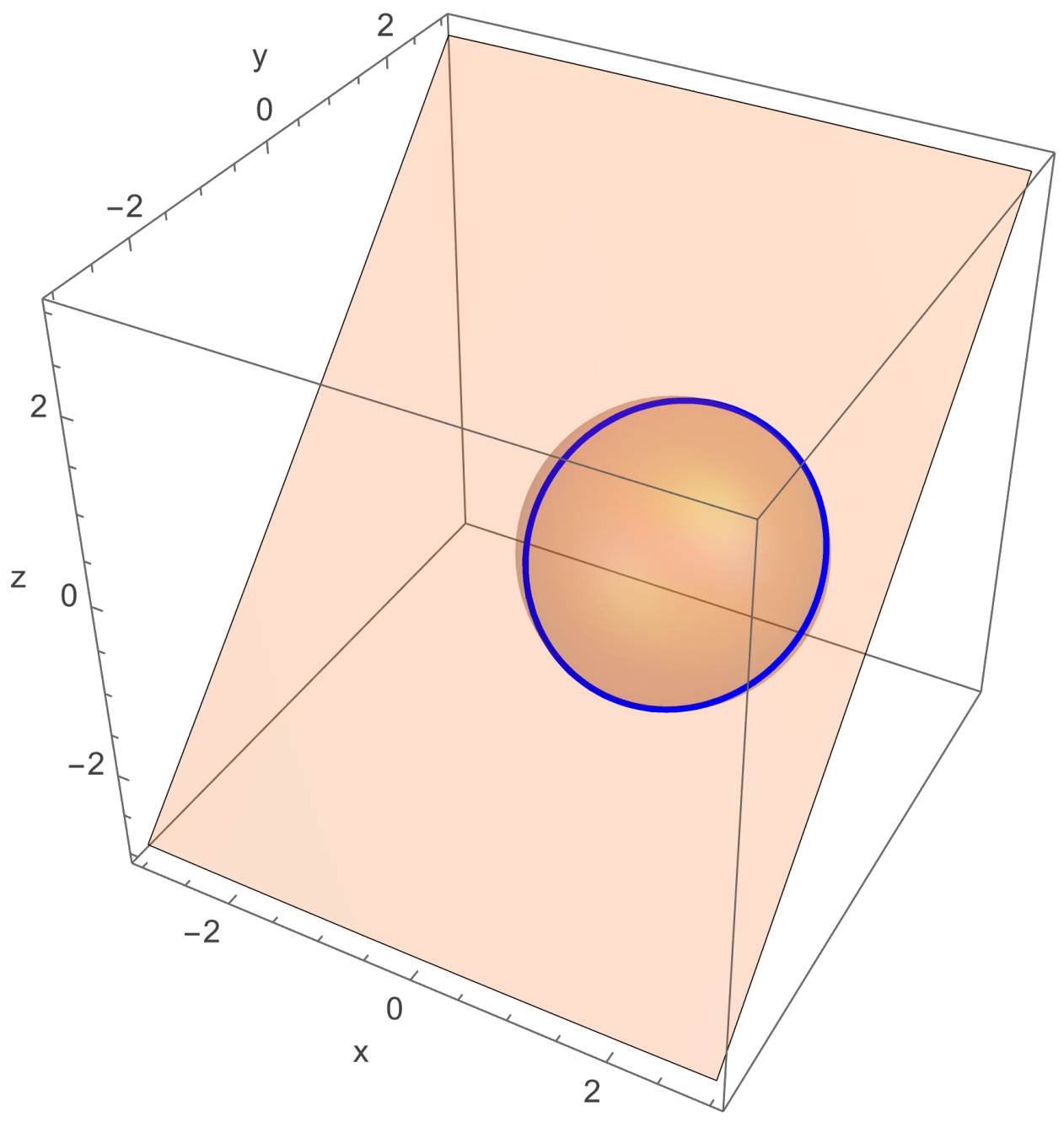

**Figure 11.** Hopf unknot formed from the intersection of the real and imaginary parts of the numerator $N(x,y,z)$.

The polynomial required in Eq. (39) is given by

$$W(x,y,\varsigma)=-1+x^2-2(x+iy)+y^2+i\varsigma/b. \tag{43}$$

The unknot focus wave mode solution in Equation (39) is given explicitly as

$$\begin{aligned} f(x,y,\varsigma,\eta)=&\frac{a_1}{a_1+i\varsigma}exp(ib\eta)\exp\left[-\frac{p^2}{4b}(a_1+i\varsigma)\right]\exp\left(-b\frac{x^2+y^2}{a_1+i\varsigma}\right)\\ &\times\left[-1-2a_1\left(\frac{x+iy}{a_1+i\varsigma}\right)+a_1^2\left(\frac{x^2+y^2}{a_1+i\varsigma}\right)+i\frac{a_1}{b}\frac{\varsigma}{a_1+i\varsigma}\right]. \end{aligned} \tag{44}$$

As mentioned earlier, this solution is nonsingular, but it contains infinite energy. A finite-energy solution is obtained by a superposition over the free parameter $b$, with the spectrum shown below

$$\psi(x,y,\varsigma,\eta)=\int_0^\infty db\, b^{\nu-1}e^{-a_2 b} f(x,y,\varsigma,\eta;b), a_2>0. \tag{45}$$

The required integrations can be carried out explicitly using the Laplace transform identity

$$\int_0^\infty db\, b^{\nu-1}\exp(-bP)\exp\left(-\frac{A}{4b}\right)=2\left(\frac{A}{4P}\right)^{\nu/2} K_\nu\left(\sqrt{AP}\right), \tag{46}$$

with the restrictions $\mathrm{Re}\{A\}>0$ and $\mathrm{Re}\{P\}>0.$ $K_\nu$ denotes the $\nu-$order of the modified Bessel function of the second kind. The intersection of the real and imaginary parts of this solution is shown in the figure below for two times.

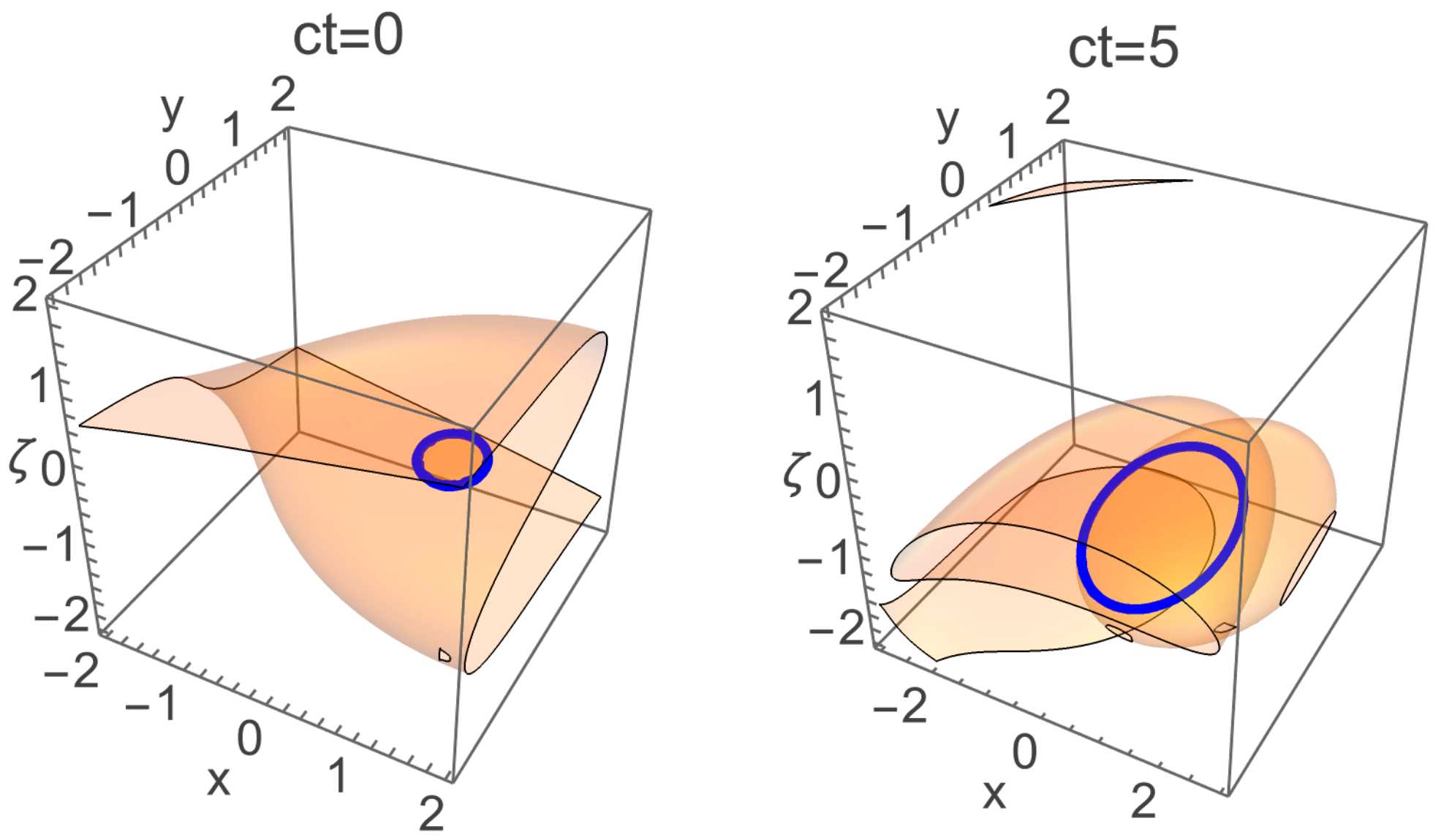

**Figure 12.** Hopf unknot formed from the intersection of the real and imaginary parts of the solution $\psi(x,y,\varsigma,\eta=\varsigma+2ct)$ at $ct=0$ and $5$. The following parameter values were used: $\nu=2, a_1=1, a_2=10$ and $p=10^{-1}$.

## 8. Discussion

In the case of knotted vector-valued electromagnetic fields (see Section 2), any two electric fields lines are linked, any two magnetic field lines are linked, and any electric field line is linked to any magnetic field line, not only simply, but also in complicated knots (Hopf, trefoil, cinquefoil, etc.) In the case of paraxial monochromatic light beams, on the other hand, one is interested in nodal knots, the latter being threads of complete darkness (lines of zero intensity) tied into diverse topological structures. Several such topological types have been successfully implemented experimentally, e.g., those arising from a superposition of Gauss-Laguerre beams [35, 36, 43].

The aim in this article has been to derive finite-energy spatiotemporally localized scalar-valued wave solutions characterized by nodal knotted topologies. This will extend to spatiotemporally localized pulsed fields the diverse important applications of monochromatic light beams, having nontrivial topological structures, in trapping and guiding particles, image processing and optical communications; also, in applications to biology and medicine.

The work in Sections 3, 4 and 5 has been confined to the basic Hopf and trefoil nodal knots. However, finite-energy spatiotemporally localized splash mode solution to the scalar wave equation with more complicated knotted nodal topologies can be easily derived.

The basic approach used Sections 3-5 has been used in Section 6 to obtain a Hopf nodal solution the {3+1)D Schrödinger equation. It has also been pointed out how to obtain, within the framework of the slowly varying envelope paraxial approximation, the same type of solution in the case of anomalous temporal dispersion by a simple change of variables. Finally, a finite-energy spatiotemporally localized splash mode solution to the Klein-Gordon equation has been derived in Section 7. Although for simplicity it is confined to an unknotted nodal topology, more complicated solutions can be easily derived.

## 9. Concluding remarks

When dealing with monochromatic light beams, not only bright regions but also those of darkness, arising from exact destructive interference, are of important physical interest, e.g., as traps for particles. Regions of zero light intensity (characterized also by phase singularities) with diverse topological structures, such as nodal knots, have been studied extensively recently. These are scalar monochromatic solutions based on the paraxial approximation. With the exception of solutions to the Schrödinger-like equation describing anomalous dispersion, no *scalar-valued spatiotemporally* localized waves with analogous knotted topological structures are known to date. In this article we have filled in this gap by deriving finite-energy spatiotemporally localized scalar-valued, knotted wave solutions in free space and in the presence of temporal dispersion.